\documentclass[aps,prb,floatfix,twocolumn]{revtex4}
\usepackage{graphicx}
\usepackage{dcolumn}
\usepackage{dcolumn}
\usepackage{amssymb}

\begin{document}

\title{Derivatives of the fixed-node energy}
\author{
S. De Palo$^{(a,b)}$,
S. Moroni$^{(a,b)}$ and S. Baroni$^{(a,c)}$ }
\affiliation{
$^a$Istituto Nazionale per la Fisica della Materia \\
$^b$Dipartimento di Fisica, Universit\`a di Roma 
``La Sapienza'' \\
$^c$Scuola Internazionale Superiore di Studi Avanzati, Trieste \\ Italy }
\gdef\theauthor{S. De Palo,
S. Moroni and S. Baroni}
\gdef\thetitle{Derivatives of the fixed-node energy}

\begin{abstract}
We present quantum Monte Carlo calculations of total energy
derivatives, consistently performed in the fixed-node
approximation. Contributions from nodal displacements, neglected or
approximated in previous investigations, are properly taken into
account.  Their impact on the efficiency is discussed.
\end{abstract}
\maketitle

\section{Introduction}

Quantum Monte Carlo simulations are emerging as a promising
alternative route towards high-accuracy predictions of molecular and
materials properties. The fixed-node approximation \cite{fnc} is
widely used in quantum Monte Carlo simulations \cite{mitas} of
many-fermion systems to avoid the sign problem \cite{kalos}.  It
consists of solving for the lowest-energy eigenstate, $\Phi$, of the
Hamiltonian, $H$, with the boundary conditions that the wave function
vanishes on the nodal surface of an anti-symmetric trial function,
$\Psi$, i.e. at the points $R$ in configuration space where
$\Psi(R)=0$. 
The fixed-node
energy (FNE)---{\em i.e.} the eigenvalue 
$E$ of the 
Hamiltonian corresponding to $\Phi$---is an upper bound for the ground
state energy, and it becomes exact if $\Psi$ has the same nodes as the
true ground state wave function, $\Phi_{ex}$. Therefore, for a given
trial function, the FNE is more accurate than the variational energy
(the expectation value of $H$ on $\Psi$), because exactness of the
latter requires the stronger condition that $\Psi$ coincide with
$\Phi_{ex}$ for all $R$.  Indeed, the fixed-node approximation offers
the best viable numerical solution for several many-fermion problems
\cite{mitas}.

The ability to calculate low-order derivatives of the
Born-Oppenheimer, energy with respect to an external parameter,
$\lambda$, is an important ingredient of the calculation of materials
properties. Forces acting on individual atoms or the stress acting
within an extended system are examples of first-order derivatives
whose vanishing determines the mechanical equilibrium of a
system. Similarly, vibrational frequencies---as well as any kind of
static generalized susceptibility---are determined by second-order
derivatives. The calculation of derivatives of the FNE, $\partial
E/\partial \lambda$, has the potential of largely extending the scope
of applications of quantum Monte Carlo simulations and spawning
further methodological developments (besides the examples given above,
$\lambda$ could also be the strength of any external field, or a
variational parameter which modifies the nodal surface of $\Psi$).
However technical difficulties \cite{hammond} have long hindered
routine calculation of FNE derivatives, and despite recent progress
\cite{zong,filippi,assaraf} open questions still remain.  In
particular, the nodal displacement upon variation of $\lambda$ has
always been neglected {\it in toto} or in part \cite{commento}, thus
adding a further bias on top of the fixed-node approximation:
referring to recent literature, $\lambda$-independent nodes have been
chosen in the path integral Monte Carlo calculation of electronic
forces \cite{zong}; they {\em have} to be chosen so in applications of
the Hellmann-Feynman theorem \cite{assaraf,flad}; finally, the
drift-diffusion factor of the propagator, together with its nodal
contributions, has been dropped from the weights in
correlated-sampling quantum Monte Carlo simulations \cite{filippi}.

In this paper we address the calculation of the derivatives of the FNE
without any further approximations. To this end, we examine in detail
a simple but significant model that, we believe, is representative of
the difficulties which are met in the calculation of these derivatives
in more realistic situations. The focus being on the role of nodal
displacements, we consider a system where such a role is crucial and a
check against exact results is possible, namely a spinless Fermi gas
in two dimensions in a weak external field.

\section{Methodological aspects and the model}

The Hamiltonian of our model system is:
\begin{equation} H=-\sum_{i=1}^N \frac{1}{2}\nabla^2_i + \lambda
\sum_{i=1}^N \cos({\bf q}\cdot{\bf r}_i), \label{H} \end{equation}
where the strength of the external field, $\lambda$, is the parameter
upon which the FNE is considered to depend, and where periodic
boundary conditions have been assumed.
The ground state is a
Slater determinant, $D$, of $\lambda$-dependent Mathieu
functions \cite{gradstein}, whereas our trial function is
$\Psi=J\times D$
with a symmetric Jastrow factor $J(R) = \exp ( -\sum_{i,j}u(r_{ij}))$
which introduces spurious pair correlations.  The fixed-node
solution, exact in this case, is recovered as an absolutely
non-trivial result of the simulation, since the trial function, wrong
on purpose, can be made poor at will by tuning the {\em
pseudo-potential}, $u$.

We use the reptation quantum Monte Carlo (RQMC) 
method \cite{rqmc},
in which the calculation of derivatives is simpler to implement than
with branching algorithms---just as simple as in path integral
Monte Carlo \cite{zong}.
The FNE can be computed from the mixed estimate of the Hamiltonian,
\begin{equation}
E= \frac{\langle\Psi|H|\Phi\rangle}{\langle\Psi|\Phi\rangle}
 = \frac{\int E_L(R) f(R) dR}{\int f(R)dR},
\label{E}      
\end{equation}
by averaging the local energy $E_L(R)=H\Psi(R)/\Psi(R)$
over the mixed distribution $f(R)=\Phi(R)\Psi(R)$.
The unknown state $\Phi$ is expressed in terms of $\Psi$ through
the imaginary-time propagator, $\Phi=\lim_{\tau\to\infty}e^{-H\tau}\Psi$.
Upon discretization of the imaginary time evolution and 
introduction of importance sampling, the mixed distribution can be cast 
into the form:
\begin{equation} f(R_M)=\lim_{M\to\infty}
\int \Pi_{i=0}^{M-1} G(R_i,R_{i+1};\epsilon)\Psi(R_0)^2, \label{f}
\end{equation}
where $\epsilon=\tau/M$ is the time step. The
importance-sampled Green's function,
$G(R,R';\epsilon)=\Psi(R)\langle R|e^{-\epsilon H} |R'\rangle/\Psi(R')$,
is replaced by a short-time approximation, a common choice being \cite{mitas}
\begin{equation}
G_F(R,R';\epsilon)
 \propto  e^{-(R'-R-\epsilon F(R)/2)^2/2\epsilon}
          e^{-\frac{\epsilon}{2}(E_L(R)+E_L(R')) },
\label{wpiu}
\end{equation} with $F=\nabla\ln\Psi^2$. 

In the RQMC method a generalized Metropolis algorithm is used to
sample the probability distribution for a path of finite length in
imaginary time (a {\em reptile}, $X=\{R_0,\ldots,R_M\}$, 
\begin{equation}
P(X)=\Pi_{i=0}^{M-1} G(R_i,R_{i+1};\epsilon)\Psi(R_0)^2, \label{P}
\end{equation}
the limit in Eq.~(\ref{f}) being reached for large enough $M$.
Replacement of Eq.~(\ref{f}) in (\ref{E}) shows that the FNE can be evaluated
by averaging the local energy on the distribution $P$,
\begin{equation}
E=\langle\langle E_L(R_M)\rangle\rangle\equiv
\frac{\int dX E_L(R_M)P(X)}{\int dX P(X)}.
\label{eave}
\end{equation}
Since $P(X)$ is explicitly known, we can differentiate to get
\begin{eqnarray}
\partial_\lambda E &  = &  \langle\langle A(X)\rangle\rangle,\\
\nonumber
A(X) &  =  & \left[  \partial_\lambda E_L(R_M)
            +(E_L(R_M)-E) \partial_\lambda \ln P(X) \right]
\label{modo_derivata}
\end{eqnarray}
Finite increments of amplitude $\Delta$ are used for the evaluation of
$\partial_\lambda E_L(R)$  
and $\partial_\lambda \ln P(X)$ on the path configurations sampled from $P(X)$.
Equivalently, finite increments can be used for $E$ via 
the correlated-sampling method:
\begin{eqnarray}
 E' &  = & \frac{\int dX P'(X) E_L'(R_M)}{\int dX P'(X)}\\
\nonumber
    &  = & \frac{\int dX P(X) W(X) E_L'(R_M)}{\int dX P(X) W(X)}, 
\label{modo_rewate}
\end{eqnarray} with $W(X) = P'(X)/P(X)$, primed quantities being
evaluated at $\lambda +\Delta$.

\begin{figure}[h]
\includegraphics[width=\columnwidth]{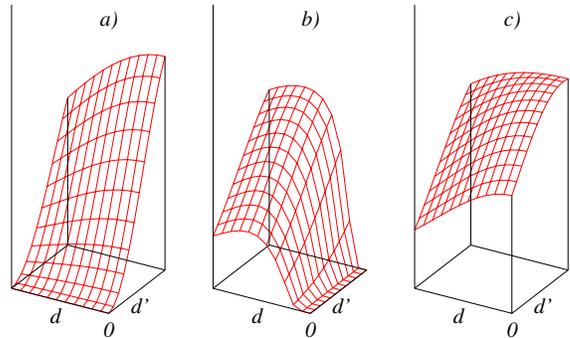}
\caption{Importance-sampled Green's function for a one-dimensional
motion near a hard wall. Left panel: exact; middle panel: Eq.~(\ref{wpiu});
right panel: Eq.~(\ref{wpiu}) with a smoothed $F$. The range of $d$ and $d'$
is $(0,\epsilon^{1/2})$.}
\label{g}
\end{figure}

In the {\it variational} Quantum Monte Carlo (VMC) approximation, the
probability distribution $P(X)$ in Eq.~(\ref{modo_derivata}) or
(\ref{modo_rewate}) is replaced by $\Psi^2(R_M)$. Small energy
differences are routinely calculated by VMC
\cite{hammond}.  The more accurate FNA requires instead the full
probability distribution of Eq~(\ref{P}), which entails nontrivial
difficulties through the term $\partial_\lambda \ln P(X)$ of
Eq.~(\ref{modo_derivata}).  Therefore we now examine in some detail
the behavior of the probability distribution $P$ when a configuration
along the path, say $R_i$, approaches a node of $\Psi$.

When the distance $d$ from the nodal surface is small enough, the
dynamics of the system is dominated by the hard wall potential due to
the fixed-node boundary conditions \cite{nota}.  The
importance-sampled Green's function should behave like that of a
one-dimensional motion close to a hard wall
$[G_0(d,d';\epsilon)\propto \frac{d'}{d}e^{-(d-d')^2/2\epsilon}
(1-e^{-2d d'/\epsilon})]$, as shown in the left panel of Fig.~(\ref{g}).
The short time approximation $G_F(R,R',\epsilon)$, instead, is a very
poor approximation near a node, because it assumes
$F=\nabla\ln\Psi^2$ to be constant during a time step, whereas $F$
diverges as $d^{-1}$. The approximation (\ref{wpiu}) for a
one-dimensional motion close to a hard wall, shown in the middle panel
of Fig.~(\ref{g}), also bears little resemblance to $G_0$.  First, $G_F$
does not vanish at $d=0$ where $G_0$ does. Sampling from $G_F$ would
thus imply an unphysical finite density at the nodes; the correct density,
which is proportional to $d^2$ can be recovered including in the
algorithm a Metropolis test based on the square of the trial function
\cite{fnc,umrigar}.  Second, $G_F$ vanishes at $d'=0$ where $G_0$ does
not.  This would imply a divergence proportional to $d^{-2}$ of $\ln
P$ at the nodes, thus making the evaluation of the derivative,
Eq.~(\ref{modo_derivata}), impossible. Such a divergence can be
eliminated, leaving the zero-time-step limit unchanged, by a suitable
cutoff on the quantum force, $\nabla\ln\Psi$. We thus replace $F$ in
Eq. \ref{wpiu} with $\bar{F}=4F(-1+\sqrt{1+F^2\epsilon/2})/(F^2\epsilon)
$ \cite{umrigar}.  The resulting Green's function, $G_{\bar F}$, is
shown in the right panel of Fig.~(\ref{g}).  With this replacement the
variance of the estimator (\ref{modo_derivata}) of the derivative is
finite.  However we empirically find that the extrapolation of the
result at $\epsilon\to 0$ can be problematic (see also Ref. [10])

In the simulation of electrons in atoms and molecules, it proves convenient
to use an interpolation between Eq.~(\ref{wpiu}) and a locally better
Green's function near the nuclei \cite{umrigar}. In the same spirit, we use
an interpolation $\tilde{G}$ between $G_{\bar F}$ and a locally better Green's 
function $G_N$ near the nodes:
\begin{equation}
{\tilde G}(R,R';\epsilon) = \alpha(d) G_{\bar F}(R,R';\epsilon)
                       + (1-\alpha(d))G_N (R,R';\epsilon)
\label{pola}
\end{equation}
where the function $\alpha(d)$ smoothly varies from 0 to 1 as $d$ goes from
0 to $\sim \epsilon^{1/2}$. In the new Green's function
\begin{eqnarray}
G_N(R,R';\epsilon) & \propto  & \frac{\Psi(R')}{\Psi(R)}e^{(R'-R)^2/4\epsilon}
(1-e^{-d' d/\epsilon})\\
\nonumber
& \times & e^{-\epsilon(V(R')+V(R))/2}
\label{is_nodalaction}
\end{eqnarray}
the potential $V$ is treated in the primitive approximation,
whereas the nodal dependence has the correct behavior of $G_0$.
Although the variance due to $\partial_\lambda \ln P(X)$ is finite using
${\tilde G}$, a time-step dependent cutoff on $d$ is
legitimate and can be useful.  

\section{Results}

We turn now to the discussion of our results for the Fermi
gas.  We first choose a very poor trial function $\Psi_I$ with 
a pseudo-potential, $u=a\exp(-br^2)$ where $a=0.9$ and $b=0.7$, yielding a 
variational energy about 20\% higher than the exact ground state energy
(whereas the FNE and the exact result coincide in this case,
since the Slater determinant is the true ground state wave function).
The derivative $\partial_\lambda E|_{x_0}$, with $\lambda_0=0.02$ and
$q=1.585$, is calculated using $\tilde{G}$. The result, shown in
Tab.~\ref{tabuno}, does indeed recover the correct value.

For comparison, we also perform calculations using
the approximation of Ref.~ \cite{filippi}, 
which drops the drift-diffusion part of
the propagator in Eq.~(\ref{modo_derivata}) and introduces an extra 
factor $\Psi^2(R_M)$ in order to recover the exact result in the limit of
perfect importance sampling. As shown in Tab.~\ref{tabuno}, the approximate
algorithm is more efficient (it gives a statistical error 3 times smaller
for the same number of path configurations). However its result for
the derivative is clearly biased.
Using a better trial function $\Psi_{II}$ with $a=0.45$
the systematic bias of the approximate algorithm becomes smaller than
statistical error.
It is interesting to note that the ratio of the statistical error
between the two methods does not seem to depend strongly on the quality
of $\Psi$.
We also have evidence that the efficiency 
depends on the number of particles only through the extra cost of sampling
configurations for a larger system
(simulations have been performed for $5,9$ and $21$ particles).

\begin{table}
\caption{ First derivative of fixed-node energy calculated with Variational
Monte Carlo (VMC), using RQMC with $\tilde G $ and using the
approximated algorithm as explained in the text}
\begin{tabular}{c c c c c c c c c }
& Exact & & VMC    & & RQMC   & &     App. Alg. & wave-function \\
\hline
&1& & 0.910(16)  & & 0.998(38) & &  0.941(14) & $\Psi_I$    \\
&1& & 0.9677(92) & & 1.038(38) & &  0.997(13) & $\Psi_{II}$ \\
\end{tabular}
\label{tabuno}
\end{table}

\begin{figure}[t]
\null\vspace{10mm}
\includegraphics[width=70mm]{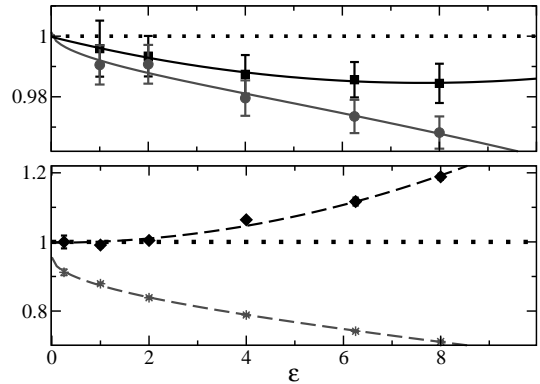}
\vspace{5mm}
\caption{Time step dependence of the first derivative 
of the FNE ($\partial_\lambda E|_{x=0.002}$), upper panel,
and of the density fluctuation ($\partial_\lambda \rho_{{\bf
q}}|_{\lambda=0}$), lower panel. We show calculations
using $\tilde{G}$ (black squares and diamonds) and 
$G_{\bar F}$ (gray circles and stars); the
black lines are quadratic fits, whereas 
the gray lines include a square root term.
\label{D1e_002}}
\end{figure}

In Fig.~\ref{D1e_002} we show the time-step dependence of various
derivatives calculated using different Green's functions.  Data
obtained with $G_{\bar F}$ have smaller statistical errors than with
$\tilde{G}$, but their extrapolation to zero time-step is harder to
guess in advance.  The upper panel shows derivatives of the FNE,
$\partial_\lambda E|_{\lambda=0.002}$. The calculation with
$\tilde{G}$ extrapolates to the exact result assuming a quadratic
time-step dependence, whereas an $\epsilon^{1/2}$ term seems necessary
in order to account for the $G_{\bar F}$ data.

The problems with the $\epsilon\rightarrow 0$ limit using $G_{\bar F}$
are more evident for the derivative of the density fluctuation,
$\partial_\lambda \rho_{{\bf q}}|_{\lambda=0}$, where $\rho_{{\bf q}}(R)=\sum_i
\exp(-i{\bf q}\cdot{\bf r}_i)$: as we can see from the lower panel of 
Fig.~(\ref{D1e_002}), in this case an even smaller power of
$\epsilon$ would be required to extrapolate to the exact result at 
$\epsilon=0$.

The physical interest in $\partial_\lambda \rho_{{\bf q}}|_{\lambda=0}$
stems from its relation to the static dielectric response. The derivative 
of $\rho_{{\bf q}}$ is evaluated as $\langle\langle B(X)\rangle\rangle$,
with $B(X)=\partial\rho_{{\bf q}}(R_{M/2})+\rho_{{\bf
q}}(R_{M/2})\partial_\lambda \ln P(X)$ (cfr. Eq.~(\ref{modo_derivata})
\cite{notaV}).  The same physical information can be formally obtained
from $\partial^2_\lambda E|_{\lambda=0}$. 
Second derivatives are in principle straightforward to evaluate, but
differentiation of Eq.~(\ref{modo_derivata}) leads to terms
containing $\partial^2_\lambda\ln P$ and $(\partial_\lambda\ln P)^2$
which have large fluctuations.

\begin{figure}[tb]
\null\vspace{10mm}
\includegraphics[width=\columnwidth]{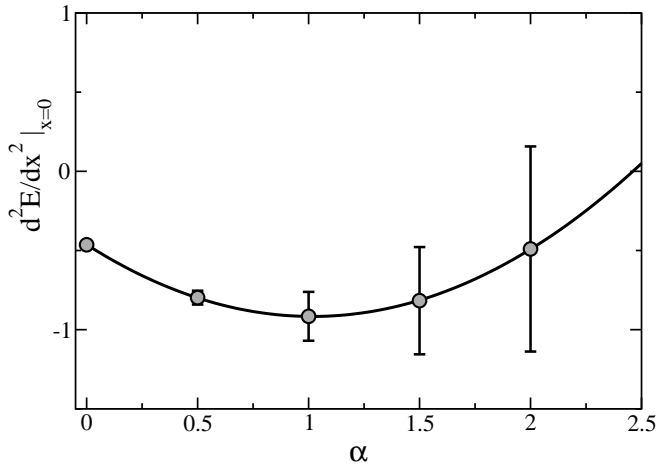}
\caption{Second derivative of the FNE, 
$\partial_\lambda^2 E|_{\lambda=0}^{(\alpha)}$ as a function of $\alpha$.
The parameter of the pseudo-potential is $a=0.90$, and the calculation
uses the Green's function $\tilde G$.
The exact result corresponds to $\alpha=1$, 
$\partial_\lambda^2 E|_{\lambda=0}^{(\alpha=1)}=-1$.}
\label{D2e_lambda}
\vspace{5mm}
\end{figure}

Despite its low efficiency, the calculation of second derivatives
of the energy offers an interesting conceptual advantage over the 
first derivative of the density fluctuation.
Suppose we don't know the exact nodal dependence of $\Psi$ on the
strength $\lambda$ of the external field. Starting from plane-wave nodes
at $\lambda=0$, we can explore a family of wave
functions, $\Psi_\alpha$, in which the one-particle orbitals are
Mathieu functions calculated for an external
potential of strength $\alpha \lambda$ {\it different} from the physical
value $\lambda$. The nodal dependence of $\Psi_\alpha$ on $\lambda$ varies
with $\alpha$: for $\alpha=0$, no nodal displacements are allowed;
for $\alpha=1$, the exact nodal dependence is recovered.
We can compute within the {\it same} run (hence in a correlated way)
the values of $\partial^2_\lambda E|_{\lambda=0}^{(\alpha)}$
pertaining to several choices of $\alpha$. The result is shown in
Fig.~(\ref{D2e_lambda}). 
Because of the variational character of the fixed node approximation, the
optimal nodal dependence (within the family $\Psi_\alpha$, and within
error bars) is given by the value $\alpha$ corresponding to the minimum of 
$\partial^2_\lambda E|_{\lambda=0}^{(\alpha)}$. Inspection of
Fig.~(\ref{D2e_lambda}) leads us to conclude that:
(i) suppression of the nodal dependence ($\alpha=0$) leads to a small
statistical error, but produces a large bias; (ii) the minimum of
$\partial^2_\lambda E|_{\lambda=0}^{(\alpha)}$ is consistent with the
exact nodal 
dependence ($\alpha=1$); (iii) the dependence of 
$\partial^2_\lambda E|_{\lambda=0}^{(\alpha)}$ on $\alpha$ is quadratic, 
so that calculations for just 3 values of $\alpha$ would suffice
(this could be useful for a higher-dimensional optimization).

Unfortunately, the calculation of second 
derivatives has still very large statistical errors (if nodal displacement
is properly included). We hope that a successful implementation of 
noise reduction techniques \cite{caffarel} could eventually turn it 
into a convenient computational tool.
Further perspectives in nodal optimization are offered by the
calculation of first derivatives of the FNE, for instance in conjunction 
with the stochastic gradient approximation method \cite{sga}.



\begin{thebibliography}{90}
\bibitem{commento} This is true with one exception \cite{vrbik};
                however it seems 
  fair to say that the results reported in Ref.~\cite{vrbik} need 
  a somewhat educated guess for the extrapolation to zero time step.
\bibitem{nota} We assume in the argument smoothness of the factor 
  $e^{-\frac{\epsilon}{2}(E_L(R)+E_L(R'))}$ of $G$.
  It is common practice to smooth the divergence of the local energy
  at the nodes when evaluating Eq.~(\ref{wpiu}).
\bibitem{notaV}	
Within RQMC it is straightforward to compute
unbiased expectation values of observables that do not commute
with the hamiltonian \cite{rqmc}.
\end{thebibliography}
\end{document}